# Validating Light Phenomena Conceptual Assessment Through The Lens of CTT and IRT Framework

Purwoko Haryadi Santoso[1,2,*], Edi Istiyono[1,3], Haryanto[1], Heri Retnawati[1]

[1]Department of Educational Research and Evaluation, Universitas Negeri Yogyakarta, Sleman 55281, Indonesia

[2]Department of Physics Education, Universitas Sulawesi Barat, Majene 91413, Indonesia

[3]Department of Physics Education, Universitas Negeri Yogyakarta, Sleman 55281, Indonesia

E-mail: purwokoharyadi.2021@student.uny.ac.id or purwokoharyadisantoso@unsulbar.ac.id

**Abstract**: Light phenomena conceptual assessment (LPCA) is a conceptual survey of light phenomena that has been recently established by the physics education research (PER) scholars. Studying the LPCA psychometric properties is always imperative to inform its measurement validity to potential LPCA users or beyond general educational researchers. Classical Test Theory (CTT) and Item Response Theory (IRT) are two popular statistical frameworks that can be utilized to explore the LPCA measurement validity. To our knowledge, no PER studies have been attempted to make a head-to-head comparison of those methods while validating the LPCA. This study was the first to delineate the LPCA measurement by statistically comparing the CTT and IRT based analysis. The LPCA dataset was drawn from physics students of eight secondary schools presented by Ndihokubwayo et al (2020). Our results accomplished the harmony between the CTT and IRT arguments to estimate the LPCA item performance and students' ability probed by the LPCA. They supported that the LPCA may be used as inventory for evaluating conceptual understanding of light phenomena from low to high students' ability range even some flagged LPCA items were exist based on CTT and IRT arguments. Special considerations for further refinement related to the discriminating power for some problematic LPCA items are discussed.

## Introduction

LPCA is a multiple-choice test recently established by Ndihokubwayo et al. (2020) to probe students' conceptual understanding of light phenomena. The LPCA comprises thirty items for 40 minutes sit that may be deployed as a conceptual inventory for secondary school through the undergraduate physics students. The LPCA has been designed to manifest the competency-based curriculum, developer experiences, and qualitative investigations with physics students. Moreover, the LPCA completes the former works concerning conceptual measurement of light phenomena established by the PER scholars such as Wave Diagnostic Test (WDT) (Wittmann et al., 1999), Light and Optics Conceptual Evaluation (LOCE) (Lakhdar et al., 2006), Light and Spectroscopy Concept Inventory (LSCI) (Bardar et al., 2009), Mechanical Waves Conceptual Survey (MWCS) (Barniol & Zavala, 2016), and Four-Tier Geometrical Optics Test (FTGOT) (Kaltakci-Gurel et al., 2017). Arguably, the LPCA is an ongoing test development process hence its refinement should follow. As such, exploring its measurement validity is always iterative for creating the more viable assessment tool (Adams & Wieman, 2011).

Using the didactic transposition theory, earlier work reported by Ndihokubwayo et al. (2020) exhibits an emphasis specifically to answer diagnostic research questions in favor of light phenomena topics, gauged by the LPCA, that are still heavily understood by their students. After the LPCA diagnoses students' thinking, their findings suggest what efforts that should be didactically managed by a physics teacher to immediately mitigate students' learning obstacles on identified light phenomena misunderstanding. To the best of our knowledge, the LPCA measurement validity has never been further assessed by Ndihokubwayo et al. (2020). This paper is the first to provide a more extended analysis of the LPCA measurement validity as a paramount step of the standard test refinement procedure established by the PER scholars (Adams & Wieman, 2011). In this study, we use a common LPCA data from Ndihokubwayo et al. (2020) whereby LPCA has been administered to secondary school physics students in pretest and posttest sessions without the learning intervention or any experimental setting.

Classical Test Theory (CTT) and Item Response Theory (IRT) have been employed in this study as the most popular analytical approaches of the multiple-choice test (Ding & Beichner, 2009). Historically, Spearman (1904) initiated the CTT paradigm at the early 20th century which proposed the observed score ($X$) can be described as the summation of true score ($T$) and measurement error ($E$) with the estimated reliability of the observed scores. The main idea behind the CTT lens is that student performance on a test is estimated based on a simply raw score. Total score can be estimated by summing up each item irrespective of the varying response patterns (Hambleton et al., 1993). Admittedly, CTT approach remains its implementation within PER to date (e.g., Brundage & Singh, 2023). Yet, it suffers certain drawbacks as a measurement paradigm (Bock & Gibbons, 2021; Prieler, 2007). CTT leads to the circular dependence of the item and the students' estimates (Baker, 2001; Bichi et al., 2015; Jabrayilov et al., 2016). Later, IRT is proposed to address potential drawbacks performed by the CTT lens (Ding & Beichner, 2009). IRT can be dating back to the law of comparative judgment from Thurstone (1927) that was extensively developed by Lord (1980) during his doctoral study. IRT is built based on the stochastic models to estimate parameters that represent the locations of examinees and items on the same latent continuum.

Bichi et al. (2015) and Setiawati et al. (2023) argues that CTT and IRT lenses may be comparable. Nonetheless, we believe that no PER scholars have made a head-to-head comparison of those competing methods while assessing the LPCA measurement validity. Admittedly, each paradigm has its pros and cons since these analytical frameworks can theoretically approach different methodologies, provide different estimates, and encompass dissimilar criteria for judging the measurement validity. To make a comparative study amongst CTT and IRT, one can propose operational criteria that should be satisfied. A common dataset has to be analyzed throughout these frameworks (Fan, 1998; Petrillo et al., 2015; Wolfaardt, 1990). To bridge a fair comparison between the CTT and IRT lenses, we chose parameters such that common psychometric properties were obtained both in terms of CTT and IRT. In this study, students' ability estimates and commonly explored item parameters (item difficulty and discrimination index) are calculated both using CTT and IRT frameworks (Baker, 2001; Bock & Gibbons, 2021;

Doran, 1980). By keeping these criteria ahead to conduct a fair comparison, the focus of this study aims to validate the LPCA psychometric properties (person and item performances) estimated by the CTT and IRT frameworks. To realize these goals, our study seeks to answer the following research questions:

1. How do the students' performance on the LPCA estimated by the CTT and IRT compared to one another?
2. How do the LPCA item parameters estimated by the CTT and IRT compared to one another?

The LPCA creators (Ndihokubwayo et al., 2020) discovered the averaged LPCA item difficulty and discrimination index are weak. Items 5 and 24 are classically discovered as the easiest items. Otherwise, items 18 and 22 are the most difficult with negative discrimination. Our study seeks to examine these results by further studying the LPCA properties based on common data disseminated by Ndihokubwayo et al. (2020). To our knowledge, our research questions intended to conduct a CTT and IRT comparison are still unknown thus far. This study offers to expand the current body of knowledge within this domain. This study can be subject to propose IRT based PER studies that are increasingly gaining favor as a modern analytical perspective of validation studies within PER community to date (Eaton et al., 2019; Richardson et al., 2021; Shoji et al., 2021). Eventually, our investigation of such comparison can recommend educators or PER researchers how to select the most appropriate way to validate conceptual physics measurement in particular on concepts covered by the LPCA.

**Method**

*Sample*

This paper was a quantitative study to validate the psychometric properties of LPCA (person and item estimates) using two competing frameworks, namely CTT and IRT. Our dataset was drawn from a previous study reported by the LPCA developers (Ndihukubwayo et al., 2020). Students in eight Rwandan secondary schools participated in the study. LPCA was given to 283 secondary school students at the pretest stage and 278 students at the posttest stage. Previous work identified 244 matched students had participated both at the pretest and posttest session. Our analyzed LPCA dataset in this paper was chosen from the LPCA pretest data. As one could argue that the larger size of pretest data can make our analysis more stable in both CTT and IRT frameworks.

We obtained the LPCA raw response data from the LPCA creators that was deposited in the Mendeley Data, an open source-based repository with Creative Commons Attribution 4.0 International License (Ndihukubwayo et al., 2020). This license allowed us to legally utilize a common dataset and we could conduct an extended LPCA validation study as long as appropriate attribution (citation) might be provided to the original LPCA creators. This data repository might be similar to PhysPort (McKagan et al., 2020), a research-based platform established by PER community. It could be used to expand the former PER studies using the shared dataset through PhysPort. Figure 1 visualizes how the LPCA dataset was downloaded from the Mendeley Data repository.

Figure 1. A snippet of the Mendeley Data platform to download the LPCA dataset.

*LPCA Measure*

The full description of the LPCA development story might be delved into Ndihokubwayo et al. (2020) paper as frequently mentioned from the introduction section. The LPCA was designed as 30 multiple-choice items with a correct answer and 3 distractors. They were deployed to probe students' conceptual understanding of light phenomena. Table 1 delineated the theoretical constructs of the LPCA explained by the LPCA creators.

Table 1. Theoretical constructs of the LPCA (Ndihokubwayo et al., 2020)

| No | Item | No | Item |
| --- | --- | --- | --- |
| 1 | Seeing | 16 | Light phenomena, fata morgana |
| 2 | Seeing | 17 | Light phenomena, sun visibility |
| 3 | Seeing | 18 | Light phenomena, sunset |
| 4 | Light phenomena, lunar eclipse | 19 | Light properties |
| 5 | Light phenomena, solar eclipse | 20 | Scattering |
| 6 | Light phenomena, the sun | 21 | Scattering |
| 7 | Reflection | 22 | Light intensity |
| 8 | Light phenomena, color of sky | 23 | Wave propagation |
| 9 | Refraction | 24 | Electromagnetic radiation |
| 10 | Refraction | 25 | Electromagnetic radiation |
| 11 | Refraction | 26 | Electromagnetic radiation |
| 12 | Light spectroscopy | 27 | Photoelectric |
| 13 | Light phenomena, rainbow | 28 | Light phenomena, the ring of sun |
| 14 | Electromagnetic spectroscopy | 29 | Wave propagation |
| 15 | Absorption spectra | 30 | Light phenomena, soap bubble |

The content validity of the LPCA was examined by students and faculty members prior to the large-scale implementation. After its content validity was assessed, they revised and piloted the LPCA at scale to 25 secondary school physics students for reliability testing using the test–retest method. The same test form was administered twice within a window of four to six weeks whereas they assumed that there was no memory effect of the students. They concluded that the LPCA has a medium degree of reliability ($\alpha$ =0.55). This value was considered acceptable given the limited number of test items and the broad range of tested knowledge (Doran, 1980). Some studies in science education acknowledge that administration of instruments genuinely testing a range of distinct knowledge facets should not be expected to give high alphas (Taber, 2018). That is, internal consistency (i.e. item equivalence) was not expected to be high in this case, because of the different physics concepts tested within the one instrument. One could argue that conceptual understanding of light phenomena may constitute a non-coherent latent construct across a multitude of students (Doran, 1980).

*Classical Test Theory (CTT)*

Prior to the IRT implementation beneath, the LPCA measurement validity was characterized by the classical perspective. We employed an open-source R based software to conduct the analysis using the "CTT" package (Sheng, 2019). This framework estimated a raw score as the student's ability estimates and some popular item performances such as item difficulty and discrimination index (Doran, 1980; Eaton et al., 2019). CTT formulated that item difficulty index was the proportion of a given sample choosing the response keyed "correct" (Doran, 1980; Hambleton et al., 1993). CTT based difficulty index was often denoted as *P* due to its operational definition as a correct "proportion" of a given sample of students. Mathematically, the item difficulty measure from the classical perspective could be presented using the equation as follows.

$$P_i = \frac{N_i}{N} \quad (1)$$

Herein, $P_i$ was the difficulty index of an item $i$, $N_i$ was the number of students who correctly answer the item $i$, and $N$ was the given sample of students taking the test. Practically, it had been suggested from Doran (1980) that viable item difficulties from CTT paradigm should be between 0.40 and 0.60, or 0.20 and 0.80. This study employed the cutoff values of 0.20 and 0.80 to justify the qualified item difficulties from CTT perspective.

The classical item discrimination index was described as a measure of how well an item discriminates the high and low students (Adams & Wieman, 2011; Doran, 1980). The simplest criteria to categorize high and low students might be taken from internal criterion, i.e., the total score on the test of which any given item is a part. Based on this criterion, the total score was ordered from high to low students. Others have suggested using the criteria of high, middle, and low students. Using this criteria, the item discrimination index could be mathematically defined using the equation (2) as follows.

$$D = \frac{\text{High (27\%)} - \text{Low(27\%)}}{N} \quad (2)$$

Herein, the middle group of students was excluded in the calculation of discrimination index. High (27%) was the item score obtained by the high ability students. Low (27%) was the item score obtained by the low ability students. Item discrimination index could be valued between 0 and 1. There were no maximum cutoff values to interpret this statistic. Yet, Doran (1980) suggested acceptable discrimination index might be between 0.2 and 0.6. This study used this limit to define the level of item discriminating power of the LPCA from the CTT paradigm.

*Item Response Theory (IRT)*

After the CTT analysis was calculated, we performed the IRT analysis on the common LPCA dataset. The IRT analysis of this study was modeled within the open-source R based software using the popular "mirt" package for IRT studies from Chalmers (2012). LPCA was a multiple-choice test such that a dichotomous response model should be applied. We built multiple dichotomous IRT models including Rasch, 1 parameter logistic (1-PL), 2-PL, 3-PL, and 4-PL model. Briefly, they were distinct based on number of parameters estimated by each model. Rasch and one-parameter model estimated merely one parameter, namely item difficulty (*b*). Two-parameter model added the item discrimination index (*a*) thus two item parameters (*a*, *b*) were estimated. Two latter IRT models employed another item parameters as represented by the number of parameters on their name. Interested readers are recommended to consult Baker (2001) for further reading about the basic of item response theory.

To decide the best fit IRT models for studying the LPCA, we examined them based on three statistical measures. The requirements of the best fit IRT model for the LPCA dataset were examined by the global fit indices such as Akaike information criterion (AIC), item fit (chi square, $\chi^2$), and person fit (Zh statistic). Table 2 introduced the reported performance of five dichotomous IRT models evaluated by all of these statistical measures.

Table 2. Global fit indices of the dichotomous IRT models on the LPCA dataset

| Model | AIC | p. $\chi^2$ | Zh |
|---|---|---|---|
| Rasch | 10431.12 | 6 infit | 1 infit |
| 1 PL | 10431.12 | 6 infit | 1 infit |
| 2 PL | 10424.97 | 4 infit | 2 infit |
| 3 PL | 10445.53 | 5 infit | 2 infit |
| 4 PL | 10467.53 | 5 infit | 1 infit |

The best fit IRT model on the LPCA data would have the lowest AIC value, thus the 2-PL model should meet. Item fit evaluated the IRT models through chi-square ($\chi^2$) statistics. It should be desirable to gain a non-significant fitness. From Table 2, we demonstrated that the number of infit items from the 2-PL model was the lowest among the others. The lack of infit items would support the 2-PL model fitness to the LPCA data (Hambleton et al., 1993). Otherwise, the measure of person fit would be examined by the Zh statistic. The small value of Zh (less than -3) was an indication of the flagged examinees (Paek & Cole, 2019). The 2-PL dichotomous model met two infit students than the other IRT models. As such, we concluded that the 2-PL was the best IRT model for studying the LPCA based on this statistical evidence. Mathematically, the 2-PL IRT model can be written as follows.

$$P_i(\theta|a_i, b_i) = \frac{1}{1 + e^{-Da_i(\theta - b_i)}} \tag{3}$$

Here, $P_i(\theta|a_i, b_i)$ was the probability of a student estimated to acquire ability $\theta$ to correctly answer item $i$ with item difficulty of $b$ and discriminating power of $a$. Then, $D$ was a constant of 1.702 which normalized the student ability continuum to correspond to a normal ogive scale more closely. The IRT difficulty index was defined as the point on the ability continuum at which the probability to correctly answer an item was 50%. The theoretical range of difficulty index was ranging between -∞ and +∞. Baker (2001) proposed the acceptable values should be between -4 and +4 as employed by this study. Furthermore, the IRT discrimination index related to an item's ability to discriminate amongst examinees. The larger an item discrimination, the better it could distinguish the high and low performing students. Functionally, discrimination index was visualized as the gradient of the item characteristic curve (ICC) of the IRT analysis. To our knowledge, even with this being the case there were no cutoff values for the IRT item discrimination index. In this study, we utilized Baker (2001) suggestion that the acceptable IRT based item discrimination index would be labeled as low at 0.35 and high at 1.35.

There were statistical prerequisites that should be satisfied to appropriately interpret the IRT based analysis. They were unidimensionality and local independence (Bock & Gibbons, 2021; Hambleton et al., 1993). Unidimensionality could be assessed based on the factor analysis. When the eigenvalue of the initial factor was more than a double of the subsequent factor (elbow method), the unidimensionality should be supported. Based on our result, this assumption was satisfied by the 2-PL model. Furthermore, one of the local independence diagnostic tests was the Yen's Q3 statistic. If our test batteries contained more than 17 items, the absolute Q3 value should be more than 0.2236. In this regard, we also invented that the 2-PL IRT model did not show any violation of the local independence assumption. After this, we explained our method to compare the LPCA person and item parameters calculated from CTT and IRT perspectives in terms of difficulty index (denoted by *P* in CTT and denoted by *b* in IRT) and discriminating power (denoted by *D* in CTT and denoted by *a* in IRT).

*CTT and IRT Comparison Method*

From the previous phases, we had computed students' ability estimates and two commonly analyzed item parameters arising from the CTT and IRT perspectives, namely difficulty and discriminating index of the LPCA items. From our research questions, the null hypothesis can be formulated that there were no significant differences of person and item parameters estimated by the CTT and IRT based analysis. To test this hypothesis, a two-tailed t-test was calculated within 95% confidence interval. It was computed across pairings of two sets of person and item parameters. Due to different scale originated from CTT and IRT estimates in nature, *z* transformation was applied to the compared person and item parameters within CTT based underlying normal variable (UNV) assumption suggested by Kohli et al. (2014).

**Results**

The results of the LPCA item analysis performed using the CTT and IRT frameworks will be discussed in this section. The presentation will follow the order of the aforementioned research questions. The estimated person estimate (students' ability) will precede, and the estimated LPCA item parameters will follow next.

CTT estimated the students' ability using the observed total score (Baker, 2001; Doran, 1980; Hambleton et al., 1993; Paek & Cole, 2019). We discovered that the mean of the students' total score was 11.2 corresponding with the standard deviation of 3.16. This value was scaled within the LPCA raw score in which it covered 30 questions in total (see Table 1). We converted it as a percentile scale or by dividing them with the total number of the LPCA items followed by a multiplication with a hundred. Through this way, we obtained the averaged CTT estimated students' ability of 37.35 and the standard deviation of 10.54. One could argue that this low performance (less than 50%) should indicate that sample of students participated in Ndihokubwayo et al. (2020) might encounter lack of understanding of light phenomena as examined by the LPCA.

IRT estimated the students' ability using the non-linear logistic model (two-parameter logistic model in our case) that calculated person and item estimates in the same continuous scale simultaneously (Baker, 2001; Bock & Gibbons, 2021). When the LPCA was analyzed using the 2-PL model, this study indicated that the mean of the students' ability was zero corresponding with the standard deviation of 0.74. This ability was scaled within the range between -4 and 4 as employed by our cutoff value of the acceptable item difficulty as explained by Baker (2001). Admittedly, this advantage of using this IRT scale was its comparability on the different samples.

Regarding the comparison task, Figure 2 displayed the density plot between CTT and IRT ability estimates. To overcome the nature of different scales within the CTT and IRT paradigms, we applied the $z$-transformation adapted from the underlying normal distribution (UNV) assumption from Kohli et al. (2014). Using Figure 2, we could immediately obtain that there was an identical distribution of students' ability shown by CTT (skewness = 0.32, kurtosis = 0.07) and IRT (skewness = 0.25, kurtosis = -0.11) estimates. Using a two-tailed $t$ test, it was found that the difference between the means was not statistically significant ($t = 8.638\text{e-}16$, $df = 564$, $p > 0.05$). This study advised CTT and IRT produced quite comparable ability estimates at the test level.

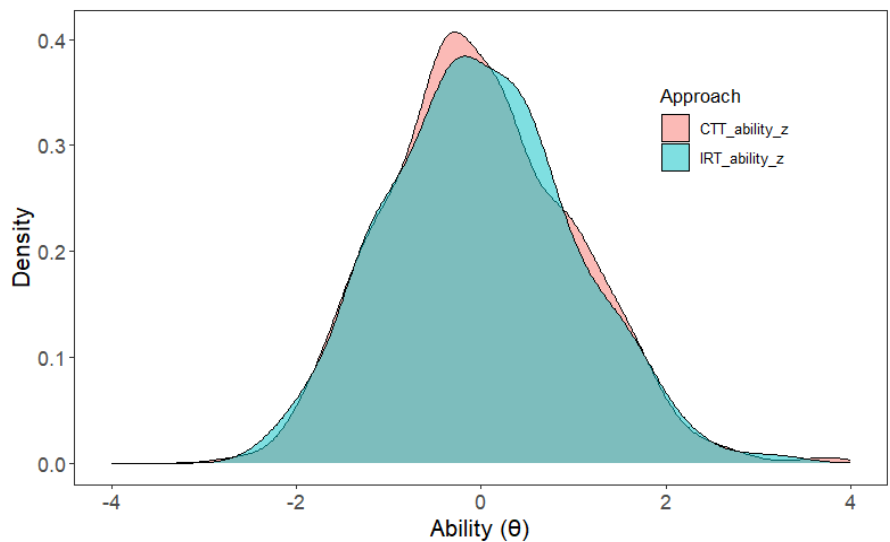

Figure 2. The distribution of the CTT and IRT based students' ability estimates.

Apart from the students' ability estimate at the test level, CTT and IRT were presumably comparable to explain the classical reliability measures. The reliability of a test was an indication of how consistency a test when it could be replicated in the other contexts (Doran, 1980; Lord, 1980). In the context of the LPCA test, we gauged the students' conceptual knowledge of light phenomena. CTT quantified the degree of the test reliability commonly using the internal consistency measured by the Cronbach' alpha (Cronbach, 1951). Reliability criteria of a good test varied with the author and purpose of the test (Taber, 2018). A value of $\alpha \geq 0.70$ was considered acceptable (Doran, 1980; Petrillo et al., 2015). Yet, in this study, we found a low level of reliability ($\alpha = 0.373$). This was lower than the value as reported in Ndihokubwayo et al. (2020) using the test-retest reliability method. One could understand this discrepancy since they calculate the LPCA reliability based on different data prior to the large-scale pretest-posttest implementation. Doran (1980, p. 104) suggested this value was considered low and therefore the LPCA should be more appropriate only for group averages or surveys.

We obtained a latent continuum of students' ability estimate analyzed by the IRT paradigm. The reliability of the LPCA in estimating students' conceptual understanding of light phenomena at the test level could be inspected through the test information function (TIF) in Figure 3. The intersection of the TIF curve (solid line, black) and the SEM curve (dashed line, red) represented the reliability of the students' abilities gauged by the LPCA test items without errors. LPCA could measure students' abilities within the range between -1.58 to 2.49. This range of students' ability was spread from the low performing students (negative theta) to the high performing students (positive theta). The greater limit of the positive theta ($\theta = 2.49$) could suggest the LPCA scale deemed to be achieved by the high ability students. According to this information curve, the LPCA performed well to accommodate a broad range of the students' understanding of light phenomena from low, medium, to high abilities. Additionally, the peak of the students' ability estimated by the IRT was 0.33. This indicated values represented by the aforementioned CTT score on the LPCA around 50% . The maximum information that could be measured by the LPCA test was 1.24. This value could be converted into a reliability estimate using the equation (reliability = 1 – [1/information]) so that the common acceptable criteria of 0.70 to 0.90 for interpreting reliability value corresponds to information of 3.3 to 1 (Kim & Feldt, 2010; Petrillo et al., 2015). Based on this formulation, our peak information $I(\theta) = 1.24$ was less than 3.3 thus the IRT reliability of the LPCA was low. This was lower than the alpha Cronbach's coefficient calculated from the CTT paradigm formerly.

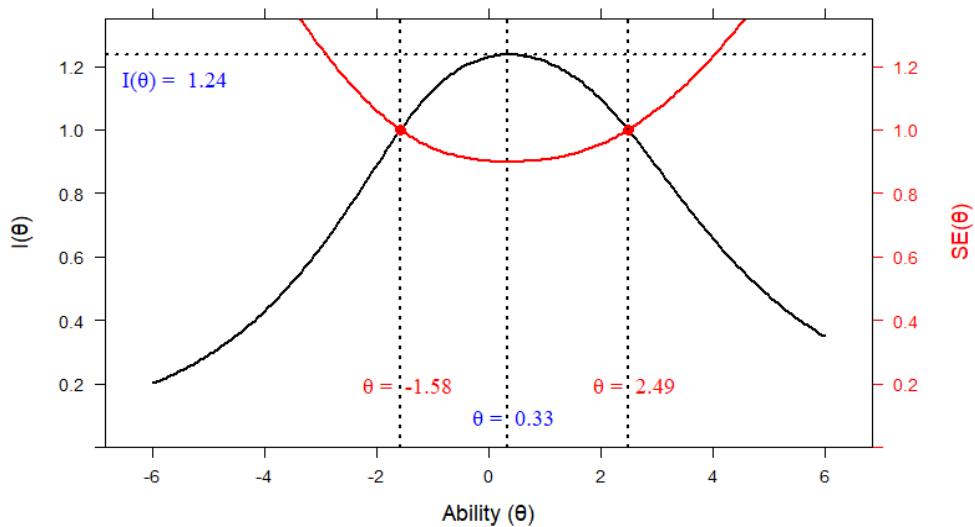

Figure 3. The TIF and SEM curves of the LPCA across students' ability ($\theta$) continuum

In Figure 4, we were interested then to report the LPCA item parameters estimated by the CTT results by inspecting a map of the item difficulty and discrimination index. Using this spatial space, the performance of the LPCA items could be identified within six graphical quadrants. The most desirable item performances should be located in quadrant II whereby the item difficulty and the discriminating index were acceptable based on criteria that have been mentioned by Doran (1980) for the CTT paradigm. We noticed that none of the LPCA items were clustered in quadrant I, III, and VI. Quadrants I and III corresponded to items with good discriminating items yet too difficult and too easy for examinee. Quadrant VI assembled the hard LPCA items with poor discriminating power. The main finding of the CTT results advised us that most of the LPCA items were on quadrant II with acceptable difficulty and acceptable for students' discrimination purpose. These results could be desirable for the LPCA creators that most of the items were qualified based on the criteria of the CTT paradigm. On the other hand, nine LPCA items were still exist on quadrant of low discriminating power. Admittedly, CTT results warranted further study intended for the LPCA test refinement in particular for enhancing its discriminating power (quadrant IV, V, VI in Figure 4). For instance, two LPCA items of quadrant IV would be further discussed below. These items were also flagged based on Ndihokubwayo et al. (2020) study.

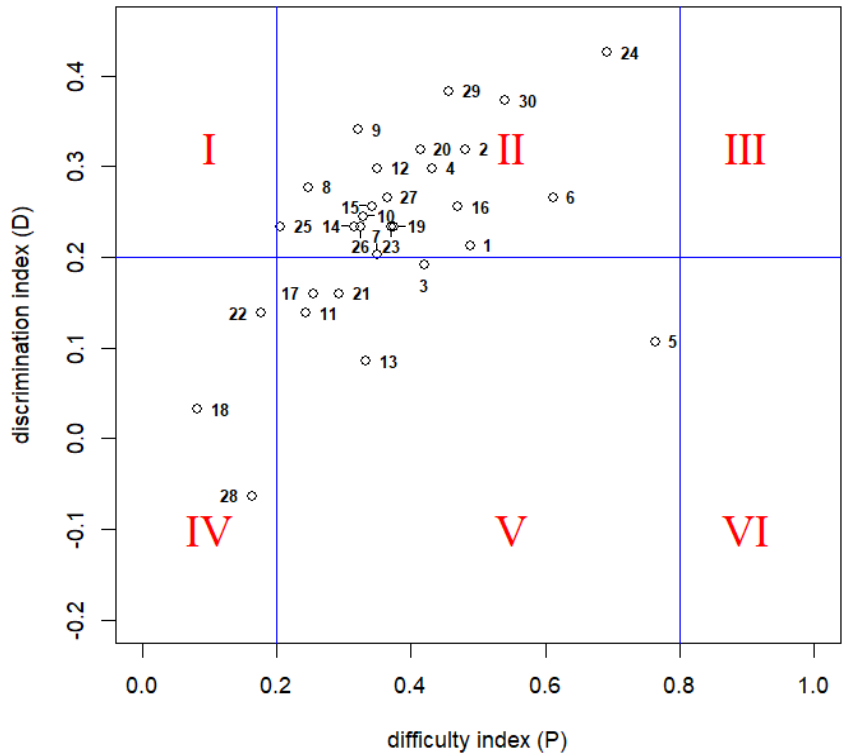

Figure 4. LPCA item difficulty and discrimination index map using CTT framework.

Using the IRT framework, we calibrated the item parameters by applying the 2-PL model to the students' data presented by Ndihokubwayo et al. (2020). On the basis of the statistical prerequisites supported by the data, 2-PL IRT model was the best fit for validating the LPCA measurement. After obtaining the support of several IRT assumptions, we studied the LPCA psychometric properties using the 2-PL IRT model. We presented Figure 5 as the IRT results in which the quadrants should be in a reverse manner than the CTT item map displayed in Figure 4 formerly. Theoretically, the scale of difficulty index was ranging between -4 and 4 as defined by the IRT model. The smaller (more negative) difficulty value corresponded to the higher probability of examinees to correctly answer the LPCA test. In the CTT case, the easier item location will be the rightmost part of Figure 4 (quadrant III and VI). To interpret the IRT results from our item map of item difficulty and discriminating power, the quadrants in Figure 5 should be reversed to maintain similar notation as delineated by the CTT results.

Quadrants I, II, III demonstrated the good discriminating LPCA items. We found that none of the LPCA items were located in quadrant III. The absence of the LPCA items in quadrant III was also agreed by the CTT results in Figure 4. There was an item in quadrant I (item 22) with a high difficulty level. Item 22 was also categorized as difficult based on CTT findings, nevertheless CTT concluded item 22 as a poor discriminating item. In fact, item 22 was in quadrant IV using the CTT lens. One could identify that there were less LPCA items clustered in quadrant II (than the CTT results) at which the most desirable item performance should be. There were 11 items in this quadrant, i.e. item 8, 9, 10, 12, 15, 20, 22, 24, 25, 26, 29, and 30. In contrast, more items (than CTT) were located in quadrant V (items 1, 2, 3, 7, 14, 16, 19, 23, 27, 28) with viable difficulty level but poor discriminating power. There were three items (18, 3, 5) congregated in quadrant VI. It might be contrasting with the CTT results (Figure 4) that we indicated items 18 and 28 placed in quadrant IV with a high difficulty level.

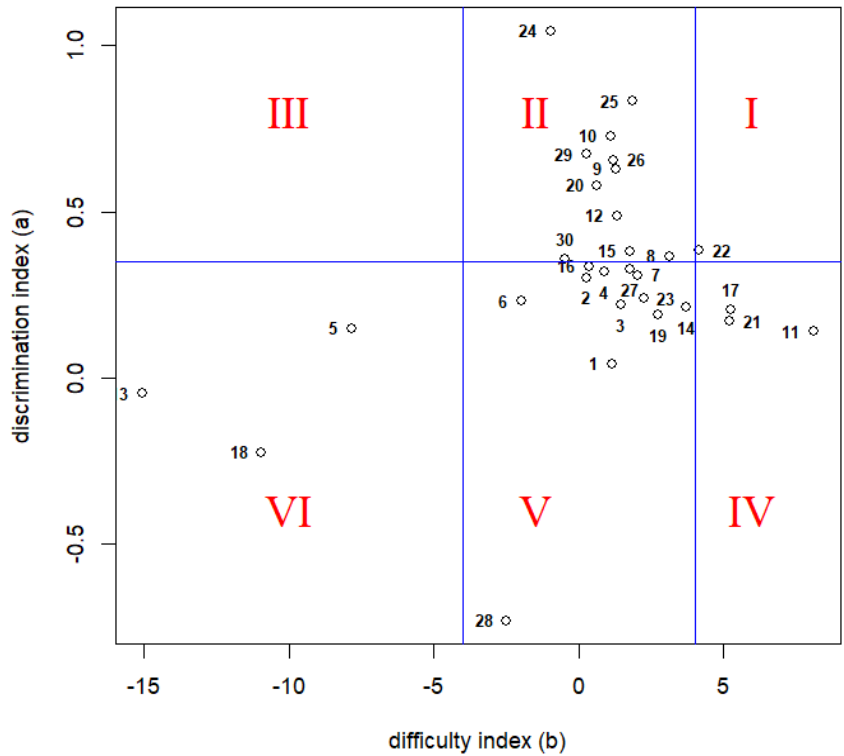

Figure 5. LPCA item difficulty and discrimination index map using IRT framework.

Table 2 summarized the shift of the LPCA item quadrants based on the CTT and IRT decisions as compared in Figure 4 and Figure 6 above. It could be interesting since one can argue that this item shift might be possible due to the different estimation approach, scaling methodology, and theoretical cutoff value proposed by the CTT and IRT paradigms. Meanwhile, our findings demonstrated that half of the LPCA items should be moved to the new quadrant after the IRT based analysis was performed. They were identified in Table 3 with an asterisk.

Item movements could be identified in Table 3. Most of them were the changes from quadrant II (in CTT) becoming quadrant V (in IRT) that might be undesirable place due to lack of discriminating power. They were well functioning in terms of the acceptable difficulty level. Secondly, we found the quadrant V (in CTT) changes to the quadrant VI (in IRT) as reflected by items 3 and 5. Thirdly, the quadrant V (in CTT) moved to the quadrant IV (in IRT) as experienced by item 11, 17, 21. The inverse of the third condition was item 28. Lastly, both CTT and IRT inferred that item 22 should be the difficult item. In contrast, CTT stated a somewhat dilemma since item 22 performed poorly discriminating power based on CTT and IRT reported the converse condition.

Table 2. The shift of the LPCA item quadrant between CTT and IRT analysis

| Item No. | CTT | IRT | Item No. | CTT | IRT | Item No. | CTT | IRT |
|---|---|---|---|---|---|---|---|---|
| 1* | II | V | 11* | V | IV | 21* | V | IV |
| 2*^ | II | V | 12 | II | II | 22* | IV | I |
| 3* | V | VI | 13 | V | V | 23*^ | II | V |
| 4*^ | II | V | 14*^ | II | V | 24 | II | II |
| 5* | V | VI | 15 | II | II | 25 | II | II |
| 6*^ | II | V | 16*^ | II | V | 26 | II | II |
| 7*^ | II | V | 17* | V | IV | 27*^ | II | V |
| 8 | II | II | 18* | IV | VI | 28* | IV | V |
| 9 | II | II | 19* | II | V | 29 | II | II |
| 10 | II | II | 20 | II | II | 30 | II | II |

*Note*: Item numbers with an asterisk were identified belong to different quadrants between the CTT and IRT results displayed in Figure 4 and Figure 5, respectively.

To explore the flagged LPCA items based on the CTT and IRT results, we were interested to shed more light on the problematic LPCA items that make shifts in Table 2 by inspecting the item characteristic curve (ICC) plot from the IRT paradigm. Only the results for items 18 and 28 were exemplified in Figure 6. We selected those items since they were categorized as flagged items in Figure 4 and 5 with low discriminatory power. Admittedly, CTT and IRT results agreed if the LPCA discriminating ability should be enhanced for further LPCA development. Figure 6 visualized the ICC plot of the items 18 and 28. These items violated the assumption of monotonicity since they produced abnormal sigmoid-shaped curve (not an "S" letter). This result could be understood that these items incorrectly estimate students' knowledge of light phenomena. Higher students were judged as lower group and conversely underperformed students were concluded as the outperformed group.

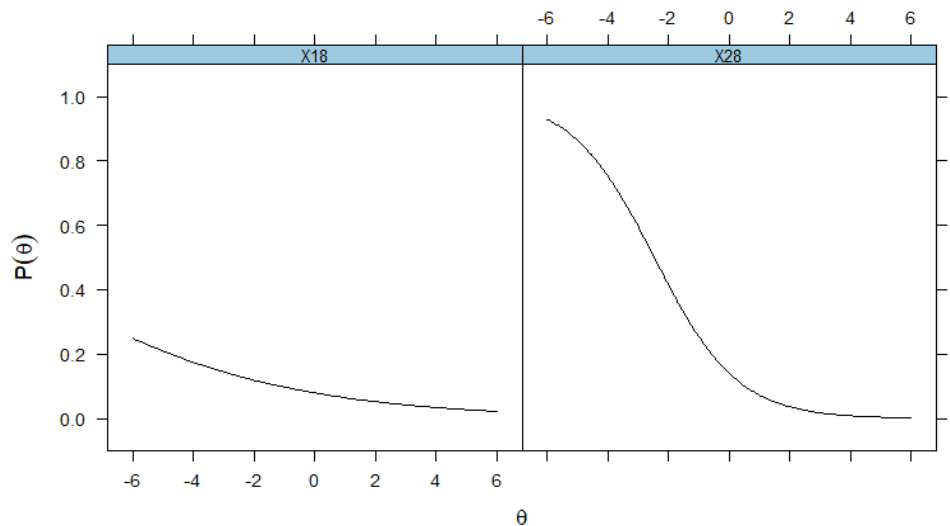

Figure 6. The ICC plot of the flagged LPCA items 18 (left) and 28 (right)

Regarding the statistical comparison of the item parameters across the CTT and IRT analysis, a two-tailed t test was performed. Due to the different manner how the CTT and IRT difficulty index should be interpreted, prior to the t-test calculation, the item difficulty index from the IRT should be reversed. There was no significant difference in terms of item difficulty index between the CTT and IRT frameworks ($t$ = -0.043673, $df$ = 58, $p$ > 0.05). Regarding the item discrimination index, we found that the difference between the means was also not statistically significant ($t$ = -1.402, $df$ = 58, $p$ > 0.05). At the item level, the CTT and IRT ability estimates reported quite comparable results on the basis of the LPCA data in this study.

**General discussion and conclusion**

This study is aimed to conduct a further validation analysis of the LPCA test through competing the lens of CTT and IRT frameworks. Instead of merely reporting the descriptive results, we seek to statistically compare the person and item estimates arising from both paradigms. In summary, we discover that there is a possible agreement between the CTT and IRT results on the basis of the LPCA data. IRT based LPCA psychometric properties confirmed the evidence that has been provided by the CTT framework to evaluate some flagged LPCA items in a more diagnostic way. Special considerations for the LPCA refinement arising from the CTT and IRT evidence should be warranted in creating the more valid and reliable LPCA measurement in theory and practice.

With regard to the first research question, this study advises that the CTT and IRT produce quite comparable ability estimates at the test level (Figure 2). This is consistent with the earlier reports of Bichi et al. (2015), Setiawati et al. (2023), Kohli et al. (2014), and Petrillo et al. (2015). Admittedly, IRT facilitates the more extended analysis rather than the CTT, for instance the plot of the test information function (TIF) (Figure 3). This plot of information can explain whether the LPCA can provide adequate information without error to measure a range of students' ability from low to high performance level. Our TIF curve shows that the thirty LPCA items can be used as a test battery for evaluating conceptual understanding of light phenomena from low to high students' abilities. We demonstrate that the LPCA has been acceptable to measure a broad range of students' knowledge level. Thus, the LPCA has been aligned as intended for the purpose of the formative assessment instruction (FASI) suggested by Adams & Wieman (2011). They argue that a FASI type instrument should be able to measure student thinking on a scale that distinguishes between novice and expert thinking. A good FASI like LPCA should gauge accurately whether students have achieved a strong, medium, or low understanding of the learned physics concept.

With regard to the second research question, the results of the item analysis underlined by the CTT and IRT lenses have been studied using a proposed map of item difficulty and discrimination index. Overall, both frameworks suggest further refinement to the LPCA items particularly improving its power to discriminate the students. The LPCA has just been developed in the PER community. Therefore, some LPCA items are still open for further development by other research. One can exemplify that two items (e.g., items 18 and 28) must be problematic based on the CTT and IRT findings. These items are assembled in quadrants with poor discriminating power that is also presented by Ndihokubwayo et al. (2020).

We have opinions regarding the LPCA test refinement, particularly for the items 18 and 28 which show a low discrimination index. The construct of stem and distractors from those is displayed in Figure 7. We may think that there is a somewhat confusing statement with the current version of the item choices. To the best of our knowledge, the conceptual explanation of the sun's redder color phenomenon in item 18 can be consulted through the literature of Young (1981) and Smith (2005). We advise that the correct explanation to the problem may be associated with the number of the atmospheric particles that sunlight must pass to our eyes. Sunlight will pass through more atmospheric molecules at the sunset than during the daytime. Particles from the atmosphere can scatter sunlight within certain wavelengths. This effect is widely known as Rayleigh scattering. When the sun sets, red is the color that scatters and reaches our eyes. According to this argument, the choice A of the item 18 should be the most appropriate answer key regarding the atmospheric particles that produce Rayleigh scattering. Furthermore, distractors A and D need to be refined since these choices can be ambiguous for the students. The terms atmosphere and sky can be understood interchangeably by the test takers. We suggest rewording of those hence they can be clearer for the test takers while choosing their answer to the item 18 of the LPCA.

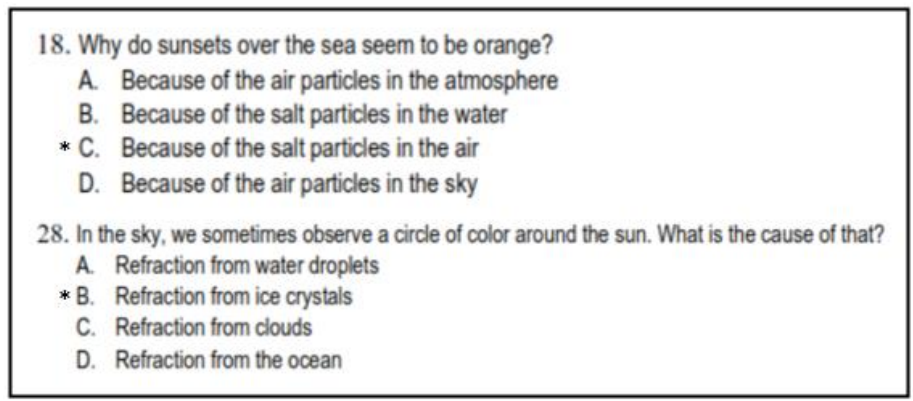

Figure 7. Item 18 and 28 of the LPCA

As described in Figure 4 and 5, the flagged LPCA items have a low discriminating power. This can be induced by some useless words as exemplified by the item 28 of the LPCA. The word ocean in the distractor D can be illogical with the examined context in the stem of the item 28. It basically questions the reason behind the colored circles around the sun or halo phenomena. Obviously, the sun may be impossible to lay on the ocean. Consequently, students receive that this distractor D should be non-sense unless they take the LPCA carelessly. The term earth's atmosphere may be more reasonable to be suggested as a distractor D. Looking at the distractor analysis, most of the students are not able to answer this item correctly. This can be an indication of the malfunctioning item. Furthermore, we recommend the conceptual term of halo phenomena needs to be mentioned in the stem. Most students may be misunderstood how the circle should be defined by the problem.

In order to enhance the LPCA distractor functions, one can suggest that they must be empirically reflected by the process of students' thinking. Adams and Wieman (2011) argue that one particularly important part of both the development and validation of a FASI type instrument is the use of student interviews to delve into the students' thinking process when they do the LPCA test. This approach is adapted from cognitive science to know to what extent students' thinking can digest the stimulus of the administered LPCA items. This practice also has been implemented by an established FASI type instrument such as Force Concept Inventory (FCI) (Stewart et al., 2018). Item distractors deployed to the FCI are based on research finding thus they can be potentially selected by the students' perspective. Admittedly, there is an open room for interested readers to attempt this recommendation in future since the phase of qualitative inquiry is still limited from Ndihokubwayo et al. (2021) study.

Our results accomplish the harmony between the CTT and IRT frameworks in terms of students' ability and item quality estimates. This study may not be construed to persuade the readers to abandon one of both frameworks in analyzing conceptual assessment. Instead, the discovered harmony may offer a more moderate standpoint to ascertain that both frameworks have their own merits to validate the conceptual measurment. The similar distribution enacted by the CTT and IRT ability estimates should be understood that sophisticated statistical methodologies proposed by the IRT paradigm maintain the similar conclusion of students' performance estimated by the CTT based observed score. Hence, high performing students inferred through the IRT analysis would also be judged at the same level by the CTT based analysis. Simplicity and less statistical prerequisites of the CTT framework should not be underestimated since it can be more intuitive for educators with little experience in sophisticated data analysis. Within the purpose of in-class formative assessment, the immediate judgment of students' performance sometimes requires physics teachers to quickly manage to help students in practice. Consequently, physics educators who work particularly in a limited time and number of students experience obstacles to use the IRT. Eventually, CTT should be more appropriate to be utilized within this circumstance.

As briefly discussed, we notice the information superiority provided by the IRT perspective can expand our validation analysis in a more diagnostic way. The same scale from the IRT between person and item estimates facilitates us to understand the LPCA performance to probe students' understanding of light phenomena. Using this latent continuum can consider the comparison of the LPCA between different contexts such as pretest-posttest data or even amongst another FASI type instrument in the same problem of interest. There is open room for further research to conduct novel study that investigates to what extent the LPCA may be correlated with the former studied light phenomena constructs established by the published FASI within PER such as LOCE (Lakhdar et al., 2006), LSCI (Barda ret al., 2009), MWCS3 (Barniol & Zavala, 2016) and FTGOT (Kaltakci-Gurel, 2017). Furthermore, the TIF curve is useful for managing item banking (Hambleton et al., 1993). The IRT offers a procedure allowing the test developer to build a test that will meet any set of desired test specifications. Thus, it is possible to build a test that discriminates well at any particular region on the ability continuum.

The choice of LPCA pretest data could be the potential limitation of the present study. It may affect the consistency of results across two psychometric methods in this paper. We believe that there is no gold standard as a basis to select the more acceptable items during the item analysis as attempted either in CTT or IRT paradigms. Item difficulty may obtain a larger attention when we are analyzing a test intended to measure students' knowledge like a formative assessment. In case of high-stakes assessment like a FASI-type instrument (Adams et al., 2011; Doran, 1980), item difficulty level may be less important, and its discrimination index should be a more emphasis. The limit of item difficulty and discrimination index defined by this study may also affect the interpretation of the CTT and IRT results. We will discover several LPCA items staying in the same quadrant for both paradigms (Table 2) when the limit of discrimination index is lower than the limit employed in this study.

Additionally, the parameter invariance assumption is one of the distinguishing person and item estimates enacted by the IRT (Bock & Gibbons, 2021; Ding & Beichner, 2009; Hambleton et al., 1993; Paek et al., 2019; Prieler, 2007). Nevertheless, this study is not intended to compare the invariant aspect performed by the CTT and IRT lenses to explore the LPCA assessment. Furthermore, one can realize that there are opportunities for other analytical perspectives with the LPCA dataset. Factor analysis should be suggested for the next research idea to investigate to what extent the LPCA is able to measure the construct of conceptual understanding on light phenomena. Through exploratory factor analysis (e.g., Santoso et al., 2022), we figure out how the LPCA measurement can be reflected by the structure of students' mental model. The multidimensional item response theory (MIRT) should be involved to accommodate the multidimensional measurement model underlied by the LPCA. This multidimensional analysis has just been carried out in the recent PER studies on several research-based assessment instruments such as Force Concept Inventory (FCI) (Stewart et al., 2018), Conceptual Survey of Electricity and Magnetism (CSEM) (Zabriskie & Stewart, 2019), Force and Motion Conceptual Evaluation (FMCE) (Yang et al., 2019), and Brief Electricity and Magnetism Assessment (BEMA) (Hansen & Stewart, 2021).

# References

Adams, W. K., & Wieman, C. E. (2011). Development and validation of instruments to measure learning of expert-like thinking. *International Journal of Science Education*, *33*(9). https://doi.org/10.1080/09500693.2010.512369

Baker, F. B. (2001). *The Basics of Item Response Theory* (2nd ed.). ERIC Clearinghouse on Assessment and Evaluation.

Bardar, E. M., Prather, E. E., Brecher, K., & Slater, T. F. (2009). Development and Validation of the Light and Spectroscopy Concept Inventory. *Astronomy Education Review*, *5*(2), 103–113. https://doi.org/10.3847/AER2006020

Barniol, P., & Zavala, G. (2016). Mechanical waves conceptual survey: Its modification and conversion to a standard multiple-choice test. *Physical Review Physics Education Research*, *12*(1). https://doi.org/10.1103/PhysRevPhysEducRes.12.010107

Bichi, A. A., Embong, R., & Mamat, M. (2015). Comparison of Classical Test Theory and Item Response Theory:A Review of Empirical Studies. *Australian Journal of Basic and Applied Sciences*, *9*(7).

Bock, R. D., & Gibbons, R. D. (2021). *Item Response Theory*. Wiley.

Brundage, M. J., & Singh, C. (2023). Development and validation of a conceptual multiple-choice survey instrument to assess student understanding of introductory thermodynamics. *Physical Review Physics Education Research*, *19*(2), 020112. https://doi.org/10.1103/PhysRevPhysEducRes.19.020112

Chalmers, R. P. (2012). mirt: A multidimensional item response theory package for the R environment. *Journal of Statistical Software*, *48*. https://doi.org/10.18637/jss.v048.i06

Cronbach, L. J. (1951). Coefficient alpha and the internal structure of tests. *Psychometrika*, *16*(3), 297–334. https://doi.org/10.1007/BF02310555/METRICS

Ding, L., & Beichner, R. (2009). Approaches to data analysis of multiple-choice questions. *Physical Review Special Topics - Physics Education Research*, *5*(2). https://doi.org/10.1103/PhysRevSTPER.5.020103

Doran, R. L. (1980). *Basic Measurement and Evaluation of Science Instruction*. National Science Teachers Association.

Eaton, P., Johnson, K., Frank, B., & Willoughby, S. (2019). Classical test theory and item response theory comparison of the brief electricity and magnetism assessment and the conceptual survey of electricity and magnetism. *Physical Review Physics Education Research*, *15*(1). https://doi.org/10.1103/PhysRevPhysEducRes.15.010102

Fan, X. (1998). Item Response Theory and Classical Test Theory: An Empirical Comparison of their Item/Person Statistics: *Educational and Psychological Measurement*, *58*(3), 357–381. https://doi.org/10.1177/0013164498058003001

Hambleton, R. K., Jones, R. W., & Rogers, H. J. (1993). Comparison of classical test theory and item response theory and their applications to test development. *Educational Measurement Issues and Practice*, *12*(3), 38–47. https://doi.org/https://doi.org/10.1111/j.1745-3992.1993.tb00543.x

Hansen, J., & Stewart, J. (2021). Multidimensional item response theory and the Brief Electricity and Magnetism Assessment. *Physical Review Physics Education Research*, *17*(2), 20139. https://doi.org/10.1103/PhysRevPhysEducRes.17.020139

Jabrayilov, R., Emons, W. H. M., & Sijtsma, K. (2016). Comparison of Classical Test Theory and Item Response Theory in Individual Change Assessment. *Applied Psychological Measurement*, *40*(8). https://doi.org/10.1177/0146621616664046

Kaltakci-Gurel, D., Eryilmaz, A., & McDermott, L. C. (2017). Development and application of a four-tier test to assess pre-service physics teachers' misconceptions about geometrical optics. *Research in Science and Technological Education*, *35*(2), 238–260. https://doi.org/10.1080/02635143.2017.1310094

Kim, S., & Feldt, L. S. (2010). The estimation of the IRT reliability coefficient and its lower and upper bounds, with comparisons to CTT reliability statistics. *Asia Pacific Education Review*, *11*(2). https://doi.org/10.1007/s12564-009-9062-8

Kohli, N., Koran, J., & Henn, L. (2014). Relationships Among Classical Test Theory and Item Response Theory Frameworks via Factor Analytic Models. *Educational and Psychological Measurement*, *75*(3), 389–405. https://doi.org/10.1177/0013164414559071

Lakhdar, Z. B., Culaba, I. B., Lakshminarayanan, V., Maquiling, J. T., Mazzolini, A., & Sokoloff, D. R. (2006). *Active learning in optics and photonics: training manual*. UNESCO. https://unesdoc.unesco.org/ark:/48223/pf0000217100

Lord, F. M. (1980). *Applications of item response theory to practical testing problems*. Routledge. https://doi.org/10.4324/9780203056615

McKagan, S. B., Strubbe, L. E., Barbato, L. J., Mason, B. A., Madsen, A. M., & Sayre, E. C. (2020). PhysPort Use and Growth: Supporting Physics Teaching with Research-based Resources Since 2011. *The Physics Teacher*, *58*(7). https://doi.org/10.1119/10.0002062

Ndihokubwayo, K., & Uwamahoro, J. (2020). Light phenomena conceptual assessment: an inventory tool for teachers. *Physics Education*. https://iopscience.iop.org/article/10.1088/1361-6552/ab6f20/meta

Paek, I., & Cole, K. (2019). *Using R for Item Response Theory Model Applications*. Routledge. https://doi.org/10.4324/9781351008167

Petrillo, J., Cano, S. J., McLeod, L. D., & Coon, C. D. (2015). Using classical test theory, item response theory, and rasch measurement theory to evaluate patient-reported outcome measures: A comparison of worked examples. *Value in Health*, *18*(1), 25–34. https://doi.org/10.1016/j.jval.2014.10.005

Prieler, J. A. (2007). So wrong for so long - Changing our approach to change. *The Psychologist*, *20*(12). https://www.bps.org.uk/psychologist/so-wrong-so-long-changing-our-approach-change

Richardson, C. J., Smith, T. I., & Walter, P. J. (2021). Replicating analyses of item response curves using data from the Force and Motion Conceptual Evaluation. *Physical Review Physics Education Research*, *17*(2). https://doi.org/10.1103/PhysRevPhysEducRes.17.020127

Santoso, P. H., Istiyono, E., & Haryanto. (2022). Principal Component Analysis and Exploratory Factor Analysis of the Mechanical Waves Conceptual Survey. *Jurnal Pengukuran Psikologi Dan Pendidikan Indonesia*, *11*(2). https://doi.org/10.15408/jp3i.v11i2.27488

Setiawati, F. A., Amelia, R. N., Sumintono, B., & Purwanta, E. (2023). Study Item Parameters of Classical and Modern Theory of Differential Aptitude Test: Is it Comparable? *European Journal of Educational Research*, *12*(2), 1097–1107. https://doi.org/10.12973/EU-JER.12.2.1097

Sheng, Y. (2019). CTT Package in R. *Measurement*, *17*(4). https://doi.org/10.1080/15366367.2019.1600839

Shoji, Y., Munejiri, S., & Kaga, E. (2021). Validity of Force Concept Inventory evaluated by students' explanations and confirmation using modified item response curve. *Physical Review Physics Education Research*, *17*(2). https://doi.org/10.1103/PhysRevPhysEducRes.17.020120

Smith, G. S. (2005). Human color vision and the unsaturated blue color of the daytime sky. *American Journal of Physics*, *73*(7), 590. https://doi.org/10.1119/1.1858479

Spearman, C. (1904). "General Intelligence," Objectively Determined and Measured. *The American Journal of Psychology*, *15*(2). https://doi.org/10.2307/1412107

Stewart, J., Zabriskie, C., Devore, S., & Stewart, G. (2018). Multidimensional item response theory and the Force Concept Inventory. *Physical Review Physics Education Research*, *14*(1). https://doi.org/10.1103/PhysRevPhysEducRes.14.010137

Taber, K. S. (2018). The Use of Cronbach's Alpha When Developing and Reporting Research Instruments in Science Education. *Research in Science Education*, *48*(6), 1273–1296. https://doi.org/10.1007/S11165-016-9602-2/TABLES/1

Thurstone, L. L. (1927). A law of comparative judgment. *Psychological Review*, *34*(4). https://doi.org/10.1037/h0070288

Wittmann, M. C., Steinberg, R. N., & Redish, E. F. (1999). Making sense of how students make sense of mechanical waves. *The Physics Teacher*, *37*(1). https://doi.org/10.1119/1.880142

Wolfaardt, J. B. (1990). Achievement Test Item Analysis: A Comparison of Traditional and Modern Methods. *South African Journal of Psychology*, *20*(4), 282–286. https://doi.org/10.1177/008124639002000408

Yang, J., Zabriskie, C., & Stewart, J. (2019). Multidimensional item response theory and the force and motion conceptual evaluation. *Physical Review Physics Education Research*, *15*(2). https://doi.org/10.1103/PhysRevPhysEducRes.15.020141

Young, A. T. (1981). Rayleigh scattering. *Applied Optics*, *20*(4), 533–535. https://doi.org/10.1364/AO.20.000533

Zabriskie, C., & Stewart, J. (2019). Multidimensional Item Response Theory and the Conceptual Survey of Electricity and Magnetism. *Physical Review Physics Education Research*, *15*(2). https://doi.org/10.1103/PhysRevPhysEducRes.15.020107